\documentclass[%
 reprint,
 onecolumn,
superscriptaddress,
 amsmath,amssymb,
 aps,
]{revtex4-2}
\usepackage{microtype}
\usepackage{graphicx}
\usepackage{hyperref}

\begin{document}
\title[Reflections on ML, automation, and materials research]{Reflections on the future of machine learning for materials research}
\date{Fall 2021}

\author{Naohiro Fujinuma}
\affiliation{Department of Chemical Engineering, Rowan University, Glassboro, NJ, USA}
\affiliation{Sekisui Chemical Co., Ltd, 2-4-4 Nishitemma, Kita-ku, Osaka, 530–8565 Japan}
\email{fujinuma@rowan.edu}
\thanks{orcid.org/0000-0002-5283-8847}

\author{Brian DeCost}
\affiliation{Material Measurement Laboratory, National Institute of Standards and Technology, Gaithersburg, MD, USA}
\email{brian.decost@nist.gov}
\thanks{orcid.org/0000-0002-3459-5888}

\author{Jason Hattrick-Simpers}
\affiliation{Department of Materials Science and Engineering, University of Toronto, Toronto, ON, Canada}
\email{jason.hattrick.simpers@utoronto.ca}
\thanks{orcid.org/0000-0003-2937-3188}

\author{Samuel E. Lofland}
\affiliation{Department of Physics and Astronomy, Rowan University, Glassboro, NJ, USA}
\email{lofland@rowan.edu}
\thanks{orcid.org/0000-0002-1024-5103}

\maketitle
\section{Abstract}
Applied machine learning (ML) has rapidly spread throughout the physical sciences; in fact, ML-based data analysis and experimental decision-making has become commonplace.
We suggest a shift in the conversation from proving that ML can be used to evaluating how to equitably and effectively implement ML for science.
We advocate a shift from a "more data, more compute" mentality to a model-oriented approach that prioritizes using machine learning to support the ecosystem of computational models and experimental measurements.
We also recommend an open conversation about dataset bias to stabilize productive research through careful model interrogation and deliberate exploitation of known biases.
Further, we encourage the community to develop ML methods that connect experiments with theoretical models to increase scientific understanding rather than incrementally optimizing materials.
Further, we encourage the community to develop machine learning methods that seek to connect experiments with theoretical models to increase scientific understanding rather than simply use them optimize materials. 
Moreover we envision a future of radical materials innovations enabled by computational creativity tools combined with online visualization and analysis tools that support active outside-the-box thinking inside the scientific knowledge feedback loop.
Finally, as a community we must acknowledge ethical issues that can arise from blindly following machine learning predictions and the issues of social equity that will arise if data, code, and computational resources are not readily available to all.

\section{Introduction}
Since Frank Rosenblatt created Perceptron to play checkers \cite{Rosenblatt1960}, machine learning (ML) applications have been used to emulate human intelligence.  The field has grown immensely with the advent of ever more powerful computers with increasingly smaller size combined with the development of robust statistical analyses. These advances allowed Deep Blue to beat Grandmaster Gary Kasparov in chess and Watson to win {\it Jeopardy!} The technology has since progressed to more practical applications such as advanced manufacturing and common tasks we now expect from our phones like image and speech recognition. The future of ML promises to obviate much of the tedium of everyday life by assuming responsibility for more and more complex processes, \textit{e.g.}, autonomous driving.

When it comes to scientific application, our perspective is that ML methods are just another component of the scientific modeling toolbox, with a somewhat different profile of representational basis, parameterization, computational complexity, and data/sample efficiency. Fully embracing this view will help the materials and chemistry communities to overcome perceived limitations and at the same time evaluate and deploy these methods with the same le   vel of rigor and introspection as any physics-based modeling methodology. Toward this end, in this essay we identify five areas in which materials researchers can clarify our thinking to enable a vibrant and productive community of scientific ML practitioners:

\begin{enumerate}
\item Maintain perspective on resources required
\item Openly assess dataset bias
\item Keep sight of the goal
\item Dream big enough for radical innovation
\item Champion an ethical and equitable research ecosystem
\end{enumerate}

\section{Maintain perspective on resources required}\label{sec:resources}

The recent high profile successes in mainstream ML applications enabled by internet-scale data and massive computation~\cite{DBLP:journals/corr/abs-2005-14165,deng2009imagenet} have spurred two lines of discussion in the materials community that are worth examining more closely. The first is an unmediated and limiting preference for large scale data and computation, under the assumption that successful machine learning is unrealistic for materials scientists with datasets that are orders of magnitude smaller than those at the forefront of the publicity surrounding deep learning. The second is a tendency to dismiss brute-force machine learning systems as unscientific. While there is some validity to both these viewpoints, there are opportunities in materials research for productive, creative ML work with small datasets and for the “go big or go home” brute-force approach.

\subsection{Molehills of data (or compute) are sometimes better than mountains}
A common sentiment in the contemporary deep learning community is that the most reliable means of improving the performance of a deep learning system is to amass ever larger datasets and apply raw computational power. This sometimes can encourage the fallacy that large scale data and computation are fundamental requirements for success with ML methods. This can lead to needlessly deploying massively overparameterized models when simpler ones may be more appropriate~\cite{d2020underspecification}, and it limits the scope of applied ML research in materials by biasing the set of problems people are willing to consider addressing. There are many examples of productive, creative ML work with small datasets in materials research that counter this notion~\cite{HattrickSimpers2018, Xue2016}.

In the small data regime, high quality data with informative features often trump excessive computational power with massive data and weakly correlated features. A promising approach is to exploit the bias-variance tradeoff by performing more rigorous feature selection or crafting a more physically motivated model form~\cite{childs2019embedding}. Alternatively, it may be wise to reduce the scope of the ML task by restricting the material design space or use ML to solve a smaller chunk of the problem at hand. ML tools for exploratory analysis with appropriate features can bring us much higher dimensional spaces even at an early stage of the research, which may be helpful to have a bird’s-eye view on our target.

There are also specific machine learning disciplines aimed at addressing the well-known issues of small datasets, dataset bias, noise, incomplete featurization, and over-generalization, and there has been some effort to develop tools to address them. Data augmentation and other regularization strategies can allow even small datasets to be treated with large deep learning models. Another common approach is transfer learning, where a proxy model is trained on a large dataset and adapted to a related task with fewer data points \cite{Yamada2019, Hoffmann2019, goetz2021addressing}.  Chen {\it et. al.} showed that multi-fidelity graph networks could be used in comparatively inexpensive low-fidelity calculations to bolster the accuracy of ML predictions for expensive high-fidelity calculations~\cite{Chen2021}. Finally, active learning methods are now being explored in many areas of materials research, where surrogate models are initialized on small datasets and updated as new data are taken with new predictions made, often in a manner that balances exploration with optimization~\cite{Lookman2019}. Generally a solid understanding of uncertainty of the data is critical for success with these strategies, but ML systems can lead us to some insights or perhaps serve as a guide for optimization which might otherwise be intractable.

We assert that the materials community would generally benefit from taking a more model-oriented approach to applied machine learning, in contrast to the popular prediction-oriented approach that many method-development papers take.  To achieve the goals of scientific discovery and knowledge generation, predictive ML must often play a supporting role within a larger ecosystem of computational models and experimental measurements. It can be productive to reassess~\cite{Bartel2020} the predictive tasks we are striving to address with ML methods; more carefully thought out applications may provide more benefit than simply collecting larger datasets and training higher capacity models.

\subsection{Writing off massive computation can lead to missed opportunities}
On the other hand, quantifying brute computation as “unscientific” can lead to missed opportunities to meaningfully accelerate and enable new kinds or scales of scientific inquiry~\cite{Holm2019}. Even without investment in massive datasets or specialized ML models, there is evidence that simply increasing the scale of computation applied can help compensate for small datasets~\cite{he2019rethinking}. In many cases, advances enabled in this way do not directly contribute to scientific discovery or development, but they absolutely change the landscape of feasible scientific research by lowering the barrier to exploration and increasing the scale and automation of data analysis. 

For example, recent advances in learned potential methods have provided paradigm-shifting performance improvements in protein structure prediction~\cite{Senior2020} and offer the potential to vastly expand the domain of atomistic material simulation. Similarly, when good physical models of data-generating processes exist, massive computation can enable new scientific applications through scalable automated data analysis systems. Recent examples include phase identification in electron backscatter diffraction (EBSD)~\cite{Kaufmann2020} and X-ray diffraction (XRD)~\cite{maffettoneCrystallographyCompanionAgent2021c}, and local structural analysis via extended x-ray absorption fine structure (EXAFS)~\cite{Timoshenko2020, Schmeide2021}. 

Even for domains where high-fidelity forward models are not available, generative models provide similar advances in data analysis capabilities. For example, a UV-Vis autoencoder trained on a large dataset of optical spectra~\cite{Stein2019} directly enabled inverse design of solid-state functional materials~\cite{Noh2019}.

 
 In light of the potential value of large-scale computation in advancing fundamental science, the materials field should make computational efficiency~\cite{DBLP:journals/corr/abs-1907-10597} an evaluation criterion alongside accuracy and reproducibility~\cite{DBLP:journals/corr/abs-2003-12206}. Comparison of competing methods using equal computational budgets can provide insight into which methodological innovations actually contribute to improved performance (as opposed to simply boosting model capacity) and can provide context for the feasibility of various methods to be deployed as online data analysis tools. Careful design and interpretation of benchmark tasks and performance measures are needed for the community to avoid chasing arbitrary targets that do not meaningfully facilitate scientific discovery and development of novel and functional materials.

\section{Openly assess dataset bias}\label{sec:bias}
\subsection{Acknowledging dataset bias}
It is widely accepted that materials datasets are distinct from the datasets used to train and validate machine learning systems for more “mainstream” applications in a number of ways.  While some of this is hyperbole, there are some genuine differences that have a large impact on the overall outlook for ML in materials research. For instance, there is a community-wide perception that all machine learning problems involve data on the scale of the classic image recognition and spam/ham problems.  While the MNIST\cite{mnist} dataset contains 280,000 labeled images, about twice the number of labeled instances in the Materials Project Database\cite{Jain2013}, other popular machine learning benchmark datasets are much more modest in size. For instance, the Iris Dataset contains only 50 samples each of three species of Iris and is treated as a standard dataset for evaluating a host of clustering and classification algorithms. As noted above dataset size is not necessarily the major hurdle for the materials science community in terms of developing and deploying ML systems; however, the data, input representation, and task must each be carefully considered.

Viewed as a monolithic dataset, the materials literature is an extremely heterogeneous multiview corpus with a significant fraction of missing entries. Even if this dataset were accessible in a coherent digital form, its diversity and deficiencies would pose substantial hurdles to its suitability for ML-driven science. Most research papers narrowly focus on a single or a small handful of material instances, address only a small subset of potentially relevant properties and characterization modalities, and often fail to adequately quantify measurement uncertainties. Perhaps most importantly, there is a strong systemic bias towards positive results~\cite{Dwan2008}. All of these factors negatively impact the generalization potential of ML systems. 

\begin{figure}[h!tbp]
    \centering
    \includegraphics[width=0.8\textwidth]{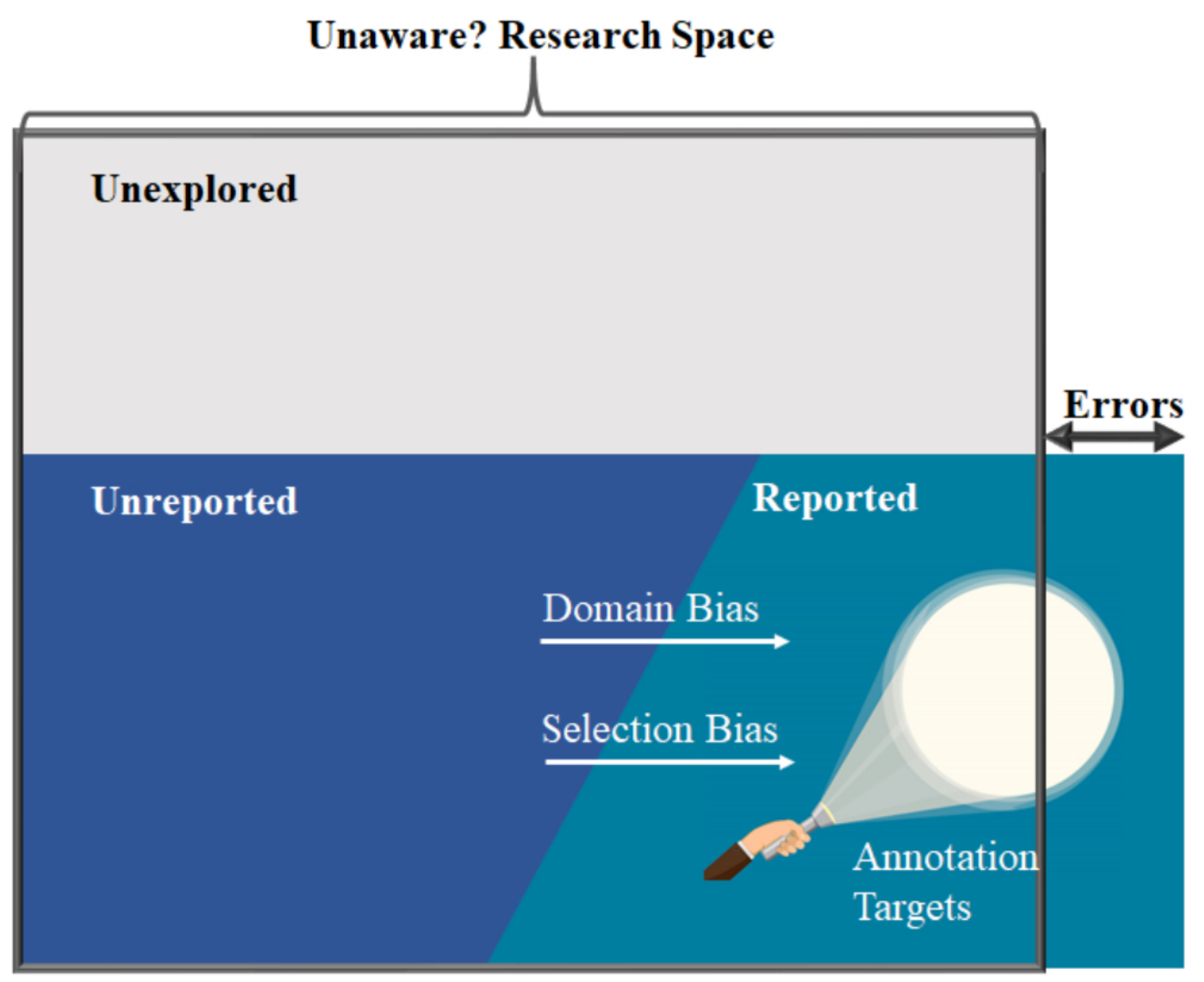}
    \caption{Where to search for new discoveries?}
    \label{fig:search}
\end{figure}

Two aspects of publication bias play a particularly large role: domain bias and selection bias. Domain bias results when training datasets do not adequately cover the input space. For example, Jia {\it et. al.} recently demonstrated that the “tried and true” method of selecting reagents following previous successes artificially constrained the range of chemical space searched, providing the AI with a distorted view of the viable parameter space~\cite{Jia2019}. Severe domain bias can lead to overly optimistic estimates of the performance of ML systems~\cite{Wallach2018, Rauer2020} or in the worst case even render them unusable for real-world scientific application \cite{griffiths2021dataset}. 

Selection bias arises when some external factor influences the likelihood of a data points inclusion in the dataset.
In scientific research, a major source of such selection bias is the large number of unreported failures.  For instance the Landolt-Bornstein collection lists 71\% of the alloys as being glass formers while the actual number of glass-forming compounds is estimated to be 5\%~\cite{10.1007/978-3-642-13850-8}.  This further complicates the already challenging task of learning from imbalanced datasets by skewing the prior probability of glass formation through dataset imbalance. Schrier {\it et. al.} reported on how incorporating failed experiments into ML models can actually improve upon the overall predictive power of a model~\cite{Raccuglia2016}.

Furthermore, the annotations or targets used to train ML systems do not necessarily represent true physical ground truth. As an example, in the field of metallic glasses the full width half-maximum (FWHM) of the strongest diffraction peak at low {\it q} is often used to categorize thin-film material as being metallic glass, nanocrystalline, or crystalline. Across the literature the FWHM value used as the threshold to distinguish between the first two classes varies from 0.4 to 0.7 \AA$^{-1}$ (with associated uncertainties) depending upon the research group. Although compendiums invariably capture the label ascribed to the samples, they almost ubiquitously omit the threshold used for the classification, the uncertainty in the measurement of the FWHM, and the associated synthesis and characterization metadata. Comprehensive studies often report only reduced summaries for the datasets presented and include full details only for a subset of “representative data.” These shortcomings are common across the primary materials science literature. Given that even experts can reasonably disagree on the interpretation of experimental results, the lack of access to primary datasets prevents detailed model critique, posing a substantial impediment to model validation~\cite{HattrickSimpers2021,griffiths2021dataset}. The push for creating F.A.I.R. (Findable, Accessible, Interoperable, and Reusable \cite{Wilkinson2016}) datasets with human/computer readable data structures notwithstanding, most of the data and meta-data for materials that have ever been made and studied have been lost to time.

 Systematic errors in datasets are not restricted to experimental results alone.  Theoretical predictions from  high throughput density functional theory (DFT) databases, for example, are a valuable resource for predicted material (meta-) stability, crystal structures, and physical properties, but DFT computations contain several underlying assumptions that are responsible for known systematic errors e.g., calculated band gaps. DFT experts are well aware of these limitations and their implications for model building; however, scientists unfamiliar with the field may not be able to reasonably draw conclusions about the potential viability of a model’s predictions given these limitations. Discrepancy between DFT and experimental data will expand as systems get increasingly more complex, a longstanding trend in applied materials science. A heterogeneous model, in particular, may cause large uncertainty depending on the complexity of the input structure, and many times little to no information is detailed about the structure or the rationale for choosing it.
 
 Finally, even balanced datasets with quantified uncertainties are not guaranteed to generate predictive models if the features used to describe the materials and/or how they are made are not sufficiently descriptive. Holistically describing the composition, structure, microstructure of existing materials is a challenging problem and the feature set used (e.g., microstructure 2-point correlation, compositional descriptors and radial distribution functions for functional materials, and calculated physical properties) is largely community driven. This presupposes that we know and can measure the relevant features during our experiments.
 Often identifying the parameters that strongly influence materials synthesis and the structural aspects highly correlated to function is a matter of scientific inquiry in and of itself.
 For example, identifying the importance of temperature in cross-linking rubber or the effect of moisture in the reproducible growth of super-dense, vertically aligned single-walled carbon nanotubes requires careful observation and lateral thinking to connect seemingly independent or unimportant variables.
 If these parameters (or covariate features, \textit{e.g.}, CVD system pump curves) are not captured from the outset, then there is no hope of algorithmically discovering a causal model, and weakly predictive models are likely to be the best case output.

\subsection{Productivity in spite of dataset bias}
Bias in historical and as-collected datasets should be acknowledged, it but does not entirely preclude their use to train an AI targeted towards scientific inquiry. Instead one can continue to gain productive insights from AI by taking the appropriate approach and thinking analytically about the results of the model.  

One method for maintaining “good” features and models is to adapt an active human intervention in the ML loop.  For example, we have recently demonstrated that Random Forest models that are tuned to aggressively maximize only cross-validation accuracy may produce low-quality, unreliable feature ranking explainability~\cite{Lei2021}. Carefully tracking which features (and data points) the model is most dependent on for its predictions allows a researcher to ensure that the model is capturing physically relevant trends, identify new potential insight into material behavior, and spot possible outliers. Similarly, when physics-based models are used to generate features and training data for ML models, subsequent comparison of new predictions to theory-based results offers the opportunity for improvement of both models~\cite{Liu2020}. An alternative approach, as recently demonstrated by Kusne {\it et. al.} is to directly have the ML model request expert input, such as performing a measurement or calculation, that is expected to lower predictive uncertainties~\cite{Kusne2020}.

Especially with small datasets, it is important to characterize the extent of dataset bias and perform careful model performance analysis to obtain realistic estimates of the generalization of ML models. See Ref.~\cite{Rauer2020} for compelling examples, an overview of recently-developed unbiasing techniques in the computational chemistry literature with details on the Asymmetric Validation Embedding method which quantifies the bias of a dataset relative to the ability of a first-nearest-neighbor model to memorize the training data. This method explicitly accounts for the label distribution but is specific to classification tasks. Leave-one-cluster-out cross-validation~\cite{Meredig2018} is more general, using only distances in input space to define cross validation groups to reduce information leakage between folds. Similarly, De Breuck {\it et. al.} used principal component analysis as a method for investigating the role of dataset bias by investigating the density of data points with scores plots~\cite{DeBreuck2021}. 

A culture of careful model criticism is also important for robust applied ML research~\cite{lipton2018troubling}. A narrow focus on benchmark tasks can lead to false incremental progress, where over time models begin overfitting to a particular test dataset and then lack generalizability beyond the initial dataset~\cite{DBLP:journals/corr/abs-1902-10811}. Recht {\it et. al.}~\cite{DBLP:journals/corr/abs-1902-10811} demonstrated that a broad range of computer vision models suffer from this effect by developing extended test sets for the CIFAR-10 and ImageNet datasets extensively used in the community for model development. This can make it difficult to reason about exactly which methodological innovations truly contribute to generalization performance. Because many aspects of ML research are empirical, carefully designed experiments are needed to separate genuine improvements from statistical effects, and care is needed to avoid {\it post-hoc} rationalization (Hypothesizing After the Results are Known (HARK)~\cite{DBLP:journals/corr/abs-1904-07633}).

That there is historical dataset bias is both unavoidable and unresolvable, but once identified this bias does not necessarily constrain the search for new materials in directions that directly contradict the bias~\cite{Nguyen2021}.  For instance, Jia {\it et. al.} identified anthropogenic biases in the design of amine-templated metal oxides, in that a small number of amine complexes had been used for a vast majority of the literature~\cite{Jia2019}.  Their solution was to perform 548 randomly generated experiments to demonstrate that a global maximum had not been reached but also to erode the systemic data bias their models observed. This is not to say that such an approach is a panacea for dataset or feature set bias as such experiments are still designed by scientists carrying their own biases (e.g., using only amines) and may suffer from uncaptured (but important!) features. Of course, a question remains how to best remove human bias from the experimental pipeline.  One might begin that endeavor by allowing researchers to use their intuition and insights for featurization, data curation, and goal setting, while permitting the ML to perform the ultimate selection of the experiment to be performed and manage data acquisition.  

\section{Keep sight of the goal}\label{sec:goal}
While the implementation of ML in materials science is goal driven, often focused on a push for better accuracy and faster calculations, these are not always the only objectives or even the most important ones.  Consider the trade-off between accuracy and discovery.  If one is optimizing the pseudopotentials to use for DFT ~\cite{Behler2007, Bartk2010}, then design is centered around accuracy. On the other hand, if the goal is to identify a material that has a novel combination of physical properties, simply knowing that such a compound exists may be sufficient to embark on a meaningful research effort.  The details related to synthesis and processing of the actual phase may likely go far beyond what is possible with any extant ML models, especially with limited benchmark datasets as one approaches the boundary of new science.  

There are clearly cases where ML is the obvious choice to accelerate research, but there can be concerns about the suitability of ML to answer the relevant question. Many applied studies focus only on physical or chemical properties of materials and often fail to include parameters relating to their fundamental utility such as reproducibility, scalability, stability, productivity, safety, or cost~\cite{olivetti2018toward}.  While humans may not be able to find correlations or patterns in high-dimensional spaces, we have rich and diverse background knowledge and heuristics; we have only just begun the difficult work of inventing ways of building this knowledge into machine learning systems.  In addition, for domains with small datasets, limited features, and a strong need for higher-level inference rather than a surrogate model, ML should not necessarily be the default approach.  A more traditional approach may be faster due to the error in the ML models associated with sample size, and heuristics can play a role even with larger datasets~\cite{George2021}. 

\begin{figure}
    \centering
    \includegraphics[width=0.8\textwidth]{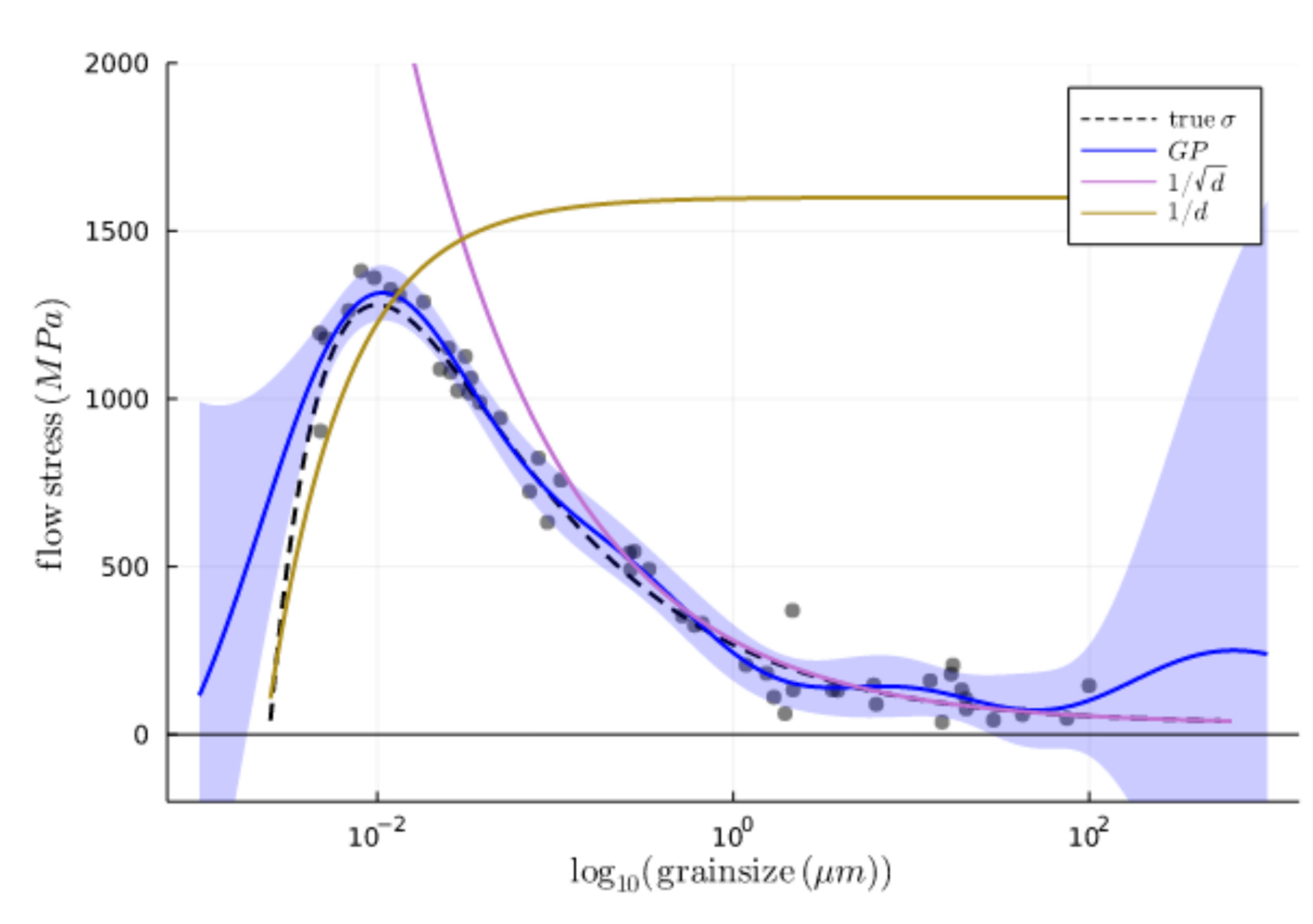}
    \caption{A Gaussian Process model can effectively reproduce the grain size dependence of the mechanical strength of an alloy even though it is completely devoid of any knowledge of the effect of the density of grain boundaries for large-grain metals \cite{Cordero2016}, the impact of grain boundary sliding in nanocrystalline alloys \cite{Trelewicz2007} or even the regime change.}
    \label{fig:hallpetch}
\end{figure}

One alternative is to employ a hybrid method which may include a Bayesian methodology to analysis~\cite{gelman1995bayesian} or may use ML to guide the work through selective intervention~\cite{hutchinson2017overcoming}. ML is only a means to model data (Figure~\ref{fig:hallpetch}), and a good fit to the dataset is no guarantee that the model will be useful since it may have little to no relationship to actual science as it attempts to emulate apparent correlations between the features and the targets.  A subsequent corollary is that any predictions from ML, especially when working with small datasets, may be unphysical.  Again, we stress that it doesn’t imply that we should never use ML for small datasets.  Rather we need to employ ML tools judiciously and understand their limitations in the context of our scientific goals.  For instance, most ML models are reasonably good at interpolation~\cite{friedman2017elements}.  On the other hand, ML is not nearly as robust when used for extrapolation, although this can be mitigated to some extent by including rigorous statistical analyses on the predictions ~\cite{Tran2020}.

A discussion of errors and failure modes can help one understand the bounds of the validity of any ML analysis although it is often lacking or limited.  An honest discourse includes not only principled estimates of model performance and detailed studies of predictive failure modes but also notes how reproducible the results within and across research groups.  Such disclosure is important for the trustworthiness of ML for any application. 

Finally, one of the biggest potential pitfalls that can occur, even for large well-curated datasets, is that one can lose sight of the goal by focusing on the accuracy of the model rather than using it to learn new science. There is a particular risk of the community spending disproportionate effort incrementally optimizing models to overfit against benchmark tasks~\cite{DBLP:journals/corr/abs-1902-10811}, which may or may not even truly represent meaningful scientific endeavors in themselves. The objective should not be to identify the one algorithm that is good at everything but rather to develop a more focused effort that addresses a specific scientific research question.   For ML to reach its true potential to transform research and not just serve as a tool to expedite materials discovery and optimization, it needs to help provide a means to connect experimental and theoretical results instead of simply serving as a convenient means to describe them. For the ML novice it is helpful to remember to keep the scientific goal at the forefront when selecting a model and designing training and validation procedures.

\section{Dream big enough for radical innovation}\label{sec:innovation}

To date, AI has increased its presence in materials science for mainly three applications: 1) automating data analysis that used to be manual, 2)serving as lead-generation in a materials screening funnel, illustrated by the Open Quantum Materials Database and Materials Project, and 3) optimizing existing materials, processes, and devices in a broadly incremental manner. While these applications are critically important in this field, we have witnessed that radical innovation historically has often been accomplished out of the context of these frameworks, driven by human interests or serendipity along with stubborn trial and error. For instance, graphene was first isolated during Friday night experiments when Geim and Novoselov would try out experimental science that was not necessarily linked to their day jobs. Escobar {\it et. al.} discovered that peeling adhesive tape can emit enough x-rays to produce images~\cite{Sanderson2008}. Shirakawa discovered a conductive polyacetylene film by accidentally mixing doping materials at a concentration a thousand times too high~\cite{Guo2020}. Design research has argued that every radical innovation investigated was done without careful analysis of a person's or even a society's needs~\cite{Norman2014}. If this is the case, an ultimate question about ML deployment in materials science would be, can ML help humans make the startling discovery of "novel" materials and eventually new science?

\begin{figure}
    \centering
    \includegraphics[width=0.8\textwidth]{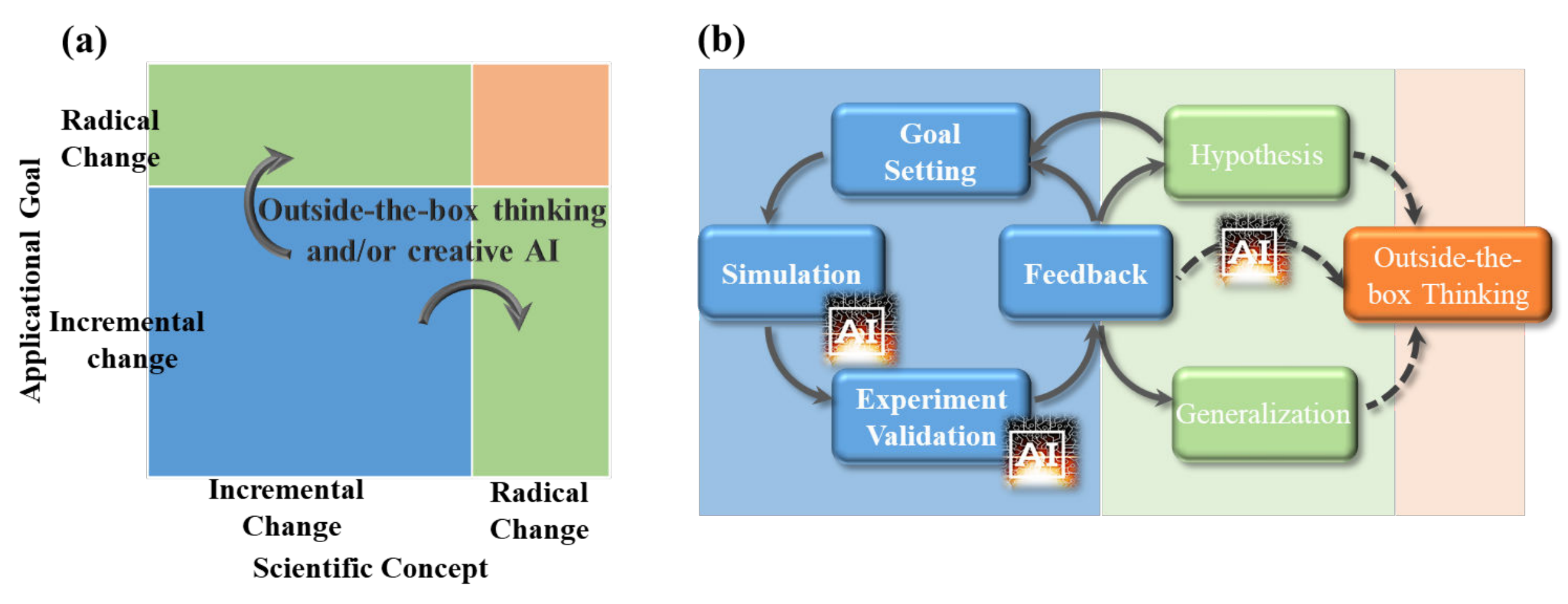}
    \caption{(a) Research categorization based upon the degree of scientific and application familiarity (b) Research loop involving machine learning with traditional and outside-the-box steps.}
    \label{fig:innovation}
\end{figure}

According to a proposed categorization in design research~\cite{Norman2014}, one can position their research based on scientific and application familiarity (Fig~\ref{fig:innovation}). Here, incremental areas (blue region) can provide easier data acquisition and interpretation of results but may hinder new discovery. In contrast, an unexplored area may more likely provide such unexpected results but presents a huge risk of wasting research resources due to the inherent uncertainty. Self-aware resource allocation and inter-area feedback will be needed to balance novelty with the probability of successful research outcomes. Although there is currently a lack of ML methods that can directly navigate one in the radical change/radical application quadrant to discover new science, we expect that there are methodologies that can harness ML to increase the chance of radical discovery.

\subsection{Active outside-the-box exploration driven by ML-assisted knowledge acquisition}
Human interests motivate outside-the-box research that may lead to a radical discovery, and these interests are fostered by theoretical or experimental knowledge acquisition. Therefore, any applied AI and automated research systems may contribute to discrete discovery by accelerating the knowledge feedback loop (Fig~\ref{fig:innovation}b). Such ML-involved research loop can include a proposal of hypotheses, theoretical and experimental examination, knowledge extraction, and generalization, which may lead to an opportunity for radical thinking. For ML to play a meaningful role in expediting this loop, one should maintain exploratory curiosity at each step and be inspired or guided by any outputs while attentively being involved in the loop. Additionally, at the very beginning of proof-of-concept research, either in a current research loop or outside-the-box search, the fear of reproducibility should not prevent the attempt at new ideas because the scientific community needs to integrate conflicting observations and ideas into a coherent theory~\cite{Redish2018}.

One can harken back to Delbruck's principle of limited sloppiness~\cite{Yaqub2018}, which reminds us that our experimental design sometimes tests unintended questions, and hidden selectivity requires attention to abnormality. In this context, ML may help us notice the anomaly or even hidden variables with a rigorous statistical procedure, leading to new pieces of knowledge and outside-the-box exploration. For instance, Nega et al. used automated experiments and statistical analysis to clarify the effect of trace water on crystal/domain growth of halide perovskite~\cite{Nega2021}, which had often been communicated only in intra-lab conversation. Since such correlation analysis can only shed light on a domain where features are input, researchers still need comprehensive experimental records containing both data and metadata to be fed, possibly regardless of their initial interests. Also, an unbiased and flexible scientific attitude based upon observation may be crucial to reconceptualize a question after finding the abnormality.

\subsection{Deep generative inverse design to assist in creating material concepts}
Functionality-oriented inverse design~\cite{Zunger2018} is an emerging approach for searching chemical spaces~\cite{Kirkpatrick2004} for small molecules and possibly solid-state compounds~\cite{ren2020inverse}. Briefly, a deep generative model learns a probabilistic latent representation of chemical space, and a surrogate model is used to optimize target properties in the latent space; novel compounds likely to have desired properties can then be sampled from the generative model~\cite{SanchezLengeling2018}. While the design spaces, such as the 166 billion molecules mapped by chemical space projects~\cite{Reymond2015}, are far beyond the human capability to understand them comprehensively, AI may distill patterns connecting functionalities and compound structures spanning the space. This approach can be a critical step in conceptualizing materials design based upon desired functionalities and further accelerating the AI-driven research loop. One application of such inverse design is to create a property-first optimization loop which includes defining a desired property, proposing a material and structure for the property, validating the results with (automated) experiments, and refining the model. 

While these generative methods may start to approach creativity, they still explicitly aim to learn an empirical distribution based on the available data. Therefore, extrapolation outside of the current distribution of known materials is not guaranteed to be productive. This suggests that these methods would probably not generate a carbon nanotube given only pre-nanotube-era structures for training or generate ordered superlattices if there is none in the training data. In addition, these huge datasets are mainly constructed based on simulation, and we need to be careful about a gap between simulated and actual experimental data as discussed previously. Still, a new concept extracted from inverse design may inspire researchers to jump into a new discrete subfield of material design by actively interpreting the abstracted property-structure relationship.

\subsection{Creative AI for materials science}
The essence of scientific creativity is the production of new ideas, questions, and connections \cite{Lehmann2019}.
The era of AI as an innovative investigator in this sense has yet to arrive.
However, since human creativity has been captured by actively learning and connecting dots highlighted by our curiosity, it may be possible that machine "learning" can be as creative as humans in order to reach radical innovation. While conventional supervised natural language processing~\cite{Krallinger2017} has required large hand-labeled datasets for training, a recent unsupervised learning study~\cite{Tshitoyan2019} indicates the possibility of extracting knowledge from literature without human intervention to identify relevant content and capturing preliminary materials science concepts such as the underlying structure of the periodic table and structure-properties relationships. This study was demonstrated by encoding latent literature into information-dense word embeddings, which recommended some materials for a specific application ahead of human discovery. Since the amount of currently existing literature is too massive for human cognition, generative AI systems may be useful to suggest a specific design or concept given appropriately defined functionalities. 

An underlying challenge is how to deal with implicit and non-machine-readable data reported in the literature. For instance, it is common to summarize experimental results with a 2D figure which just describes some tendency in a limited range along with some maxima/minima. Such disproportionate summarization does not span the entire range of the experimental space described in the figure, and may bias the parameter space that a model might explore depending upon how the literature is written. This also returns us to the issue of addressing the hesitancy of publishing “unsuccessful” research data. One may need to be careful in accepting AI-driven proposals since there is likely a gap between a human-interest-driven leap and a ML-driven suggestion based on some learned representation of the unstructured data gleaned from the literature.

Beyond latent variable optimization, one may consider computational creativity, which is used to model imagination in fields such as the arts~\cite{DBLP:journals/corr/abs-2006-08381}, music~\cite{DBLP:journals/corr/abs-1709-01620}, and gaming. This endeavor may start with finding a vector space to measure novelty as a distance~\cite{berns2020bridging}. A novelty-oriented algorithm searches the space for a set of distant new objects that is as diverse as possible as to maximize novelty instead of an objective function~\cite{lehman2011abandoning}. Since there would be some bias for measuring the distance along with exploratory space,  deep learning novelty explorer (DeLeNox) was recently proposed~\cite{DBLP:journals/corr/abs-2103-11715} as a means to dynamically change the distance functions for improved diversity. These approaches could be applied to materials science to diversify research directions and help us pose and consider novel materials and ideas though measuring novelty may be subjective and most challenging for the community, and one always needs to be mindful of ethical and physical materials constraints.

\section{Champion an ethical and equitable research ecosystem}\label{sec:ethics}
Looking toward the future of the use of ML in materials science, there are issues, such as potential physical, economic, and legal risks, that have yet to be fully discussed and resolved.
For example, ML may predict mixing several materials together to form a new compound with a set of desired properties, but the synthesis is dangerous because of the toxic gases produced during a side reaction or the final product is flammable or explosive.
Also consider that indiscriminate use of ML could lead to infringement upon intellectual property rights if the algorithm is unaware of the protected status of certain processes or materials.
A yet unanswered question regarding either scenario is, who is the responsible party - the person who created the ML environment or the person who provided the data which did not capture all potential hazards and conflicts?  It is paramount that the community reach a consensus on issues such as this before widespread autonomous use of ML.

Another concern to be addressed as ML transforms materials research is the prospect for enormous inequities between the computationally rich and poor, where the rich quickly explore large parameter spaces and the have-nots fall behind, unable to compete. This disparity would grow larger and faster if end users, reviewers, and program managers deem that only resource-intensive ML is trustworthy.  Although a materials cloud platform~\cite{klimeck2008nanohub, Talirz2020} could help to bridge the gap between these groups, it would be meaningless without a strong culture of open publication of training source code, model parameters, and appropriate benchmark datasets. Yet even making these resources freely available may still be insufficient to sustain a level playing field unless there is equivalent access to state-of-the art instrumentation to validate the increasingly more detailed predictions. Clearly, we have time before we arrive at that reckoning, but the complexity of the matter requires us to begin discussing it now.

\section{Summary}
Machine learning has been effective at expediting a variety of tasks, and the initial stage of its implementation for materials research has already confirmed that it has great promise to accelerate science and discovery~\cite{baker2019workshop}.  To realize that full potential, we need to tailor its usage to answer well defined questions while keeping perspective of the limits of the resources needed and the bounds of meaningful interpretation of the resulting analyses.  Eventually, we may be able develop ML algorithms that will consistently lead us to new breakthroughs in an open and equitable framework.  In the meantime, a complementary team of humans, AI, and robots has already begun to advance materials science for the common good.

\bibliography{mlmat}
\end{document}